\documentclass[aps,pre,reprint,twocolumn,superscriptaddress,longbibliography,floatfix]{revtex4-2}

\usepackage[colorlinks,linkcolor=blue,citecolor=blue,anchorcolor=blue,urlcolor=blue]{hyperref}
\usepackage{graphicx}
\usepackage{amsmath,amssymb}
\usepackage{siunitx}

\newcommand{\Prob}{\mathop{\mathrm{Prob}}}

\begin{document}

\title{Intersection Exponents of Simple Random Walks in Two and Three Dimensions}

\author{Qiyuan Shi}
\affiliation{Department of Modern Physics, University of Science and Technology of China, Hefei, Anhui 230026, China}
\affiliation{Hefei National Laboratory, University of Science and Technology of China, Hefei, Anhui 230088, China}

\author{Runsheng Liu}
\affiliation{School of Mathematical Sciences, Peking University, Beijing 100871, China}

\author{Xinyi Li}
\email{xinyili@bicmr.pku.edu.cn}
\affiliation{Beijing International Center for Mathematical Research, Peking University, Beijing 100871, China}

\author{Ming Li}
\email{lim@hfut.edu.cn}
\affiliation{School of Physics, Hefei University of Technology, Hefei, Anhui 230009, China}

\author{Youjin Deng}
\email{yjdeng@ustc.edu.cn}
\affiliation{Department of Modern Physics, University of Science and Technology of China, Hefei, Anhui 230026, China}
\affiliation{Hefei National Laboratory, University of Science and Technology of China, Hefei, Anhui 230088, China}
\affiliation{Hefei National Research Center for Physical Sciences at the Microscale, University of Science and Technology of China, Hefei, Anhui 230026, China}

\begin{abstract}
The probability that several independent random walks avoid one another decays algebraically with the number of steps $N\to\infty$ in two and three dimensions, $P_N\sim N^{-\xi/2}$,
where the intersection exponent $\xi$ depends on the number of random walks.
More generally, one may consider several groups of independent random walks,
with intersections allowed within each group and forbidden between
different groups.
We first study intersection exponents in two dimensions, where our results
agree with the known exact formulas and test the numerical approach.
In three dimensions, where no general exact expression is known, we determine $\xi$ for a range of cases. For two groups containing $k$ and $m$ random walks, respectively, we obtain the exponents $\xi(k,m)$ for $k=1,2$ and $m=1,2,\ldots,7$.
We further extend $m$ to a continuous parameter $\lambda$, allowing us to
obtain numerical values for $\xi(k,\lambda)$ for $k=1,2,3$ and
$0.25\leq\lambda\leq3$.
Away from the smallest moment orders, these functions increase with $\lambda$ with decreasing slopes, as expected from the strict concavity of Brownian intersection exponents. We also investigate three-group configurations for several representative cases. The resulting integer and continuous exponents supply numerical values for analytical studies of non-intersecting random paths.
\end{abstract}

\maketitle

\section{Introduction}
\label{sec:intro}

Random walks were introduced by Pearson in 1905 \cite{Pearson1905}.  In the same year, Einstein related the irregular motion of suspended particles to molecular diffusion \cite{Einstein1905}.  A random walk represents a trajectory as a succession of random steps \cite{LawlerLimic2010}.  On a regular lattice, the walker moves at each step from its current site to one of the nearest-neighbor sites.  This simple model describes charge transport in amorphous solids \cite{ScherMontroll1975}, diffusion in porous materials and biological tissues \cite{NovikovEtAl2011,AlemanyEtAl2022}, intracellular search processes \cite{BenichouEtAl2011}, and cell migration in structured environments \cite{PlankSimpsonBaker2025,CarlinEtAl2025}.  It is also used for animal foraging \cite{BenichouEtAl2011,ChupeauEtAl2015} and transport on complex networks \cite{MasudaEtAl2017}.
Across these applications, random-walk trajectories exhibit a variety of geometric and dynamical properties, which can be characterized by quantities such as displacement, first-passage times, the number of distinct sites visited, and the local times accumulated at individual sites \cite{MontrollWeiss1965,Redner2001,BenichouEtAl2011,HaoLiOkadaZheng2025}.

These observables are also basic quantities in the statistical physics of random walks \cite{Chandrasekhar1943,MontrollWeiss1965,Redner2001}. Consider a simple lattice walk $S(t)\in\mathbb Z^d$ followed for $N$ steps in $d$ spatial dimensions. Green-function and generating-function methods determine its propagator, and at large $N$ the position distribution approaches a Gaussian \cite{MontrollWeiss1965}.  The typical distance $R$ from the starting point then grows as $R\sim N^{1/2}$ \cite{Chandrasekhar1943,MontrollWeiss1965}.  First-passage and survival distributions describe when the path reaches a prescribed site or boundary and set the time scales of diffusion-controlled reactions and random searches \cite{Redner2001,BrayMajumdarSchehr2013}.

The complete trajectory is also characterized by the region it visits.  For a walk of $N$ steps, the set of all lattice sites visited from the starting point through step $N$ is denoted by
\begin{equation}
 S[0,N]=\{S(t):0\leq t\leq N\}.
 \label{eq:visited-set}
\end{equation}
Here $t=0,\ldots,N$ is the discrete step number. The values $t=0$ and $t=N$ denote the starting point and the endpoint after $N$ steps, respectively. The mean number of distinct sites visited up to step $N$ scales as \cite{DvoretzkyErdos1951}
\begin{equation}
    \left\langle\left|S[0,N]\right|\right\rangle \propto \begin{cases}
        N^{1/2}, & d=1, \\
        N/\ln N, & d=2, \\
        N, & d\geq3.
    \end{cases}
    \label{eq:range-scaling}
\end{equation}
The dimension-dependent growth of $\left|S[0,N]\right|$ reflects the extent to which the walk revisits sites along its trajectory. In lower dimensions, recurrent returns cause the path to retrace a substantial part of its previous trajectory, whereas in higher dimensions the walk progressively explores new sites.
When several independent walks are present, their visited sets may contain common lattice sites.  These common sites are intersections of the trajectories \cite{CardyTauber1996,Rast2022}.

Let $S_1$ and $S_2$ be independent simple random walks in $\mathbb Z^d$, both starting from the origin, and use $S_i[a,b]$ for the corresponding visited set of walk $i$ from step $a$ through step $b$. Removing the unavoidable intersection at $t=0$, the non-intersection probability is
\begin{equation}
 P_N(1,1)=\Prob\!\left(S_1[1,N]\cap S_2[0,N]=\varnothing\right).
 \label{eq:nonintersection-probability}
\end{equation}
For two groups containing $k$ and $m$ walks, respectively, we define $P_N(k,m)$ as the probability that no walk in one group intersects any walk in the other group, while intersections among walks within the same group are allowed. More generally, $P_N(n_1,\ldots,n_p)$ denotes the probability for $p$ groups, where the $i$th group contains $n_i$ walks. Walks belonging to different groups are required to remain mutually disjoint, while intersections within each group are allowed. Since walks belonging to different groups would otherwise intersect at their common starting point, the common initial point is excluded from the intersection condition. In simulations, walks from different groups can equivalently be initialized at distinct sites within a finite region of fixed size; such a microscopic change affects only the nonuniversal amplitude of $P_N$, not its asymptotic behavior with $N$.

The non-intersection probability can be interpreted as the survival probability of diffusing particles that are removed when trajectories from different groups meet. A meeting event then represents either a reaction or the end of a surviving history \cite{Redner2001,BrayMajumdarSchehr2013,CardyTauber1996}. Path intersections also enter random-path representations of lattice spin and scalar-field models.  In these representations,
intersections between paths play an important role in controlling connected correlations and their bounds
\cite{Symanzik1966,Symanzik1969,BrydgesFrohlichSpencer1982,FernandezFrohlichSokal1992}.  The decay of $P_N$ measures the large-scale cost of suppressing such encounters.

It is well known that the typical displacement of a random walk grows as
$R\sim N^{1/2}$, corresponding to a path fractal dimension $d_f=2$ for
$d\geq2$. This dimensionality gives a critical dimension $d_c=4$ for the
intersection of random-walk paths \cite{DvoretzkyErdosKakutani1950,ErdosTaylor1960,Lawler1982FourDimensions,
Lawler1985FourDimensions,Duplantier1988CMP}.
Below this dimension, the embedding space does not provide sufficient
freedom for two infinitely extended paths of dimension $2$ to avoid one another, and the
paths intersect with probability one as $N\to\infty$. Above it, the
higher-dimensional space allows the paths to avoid one another, leaving a
finite probability of non-intersection even as $N\to\infty$
\cite{DvoretzkyErdosKakutani1950,ErdosTaylor1960}. In the cases $d=2$ and $3$, the non-intersection probability $P_N$ vanishes as $N\to\infty$ and has the asymptotic form
\cite{BurdzyLawler1990,LawlerPuckette2000}
\begin{equation}
    P_N(n_1,\ldots,n_p)\sim N^{-\xi/2},
    \label{eq:intersection-definition}
\end{equation}
where $\xi$ is the intersection exponent, which depends on the numbers of walks in each group and on the spatial dimension.

In two dimensions, the intersection exponents can be predicted using conformal field theory and have been rigorously established in the framework of Schramm-Loewner evolution
\cite{DuplantierKwon1988,LSW2001a,LSW2001b,LSW2002twoSided},
\begin{equation}
    \xi = \frac{1}{48} \left[ \left( \sum_{i=1}^{p}\sqrt{24n_i+1}-p \right)^2-4 \right],
    \label{eq:exact-2d-packets}
\end{equation}
where $p$ is the number of groups and $n_1,\ldots,n_p$ are the corresponding group sizes. For $p=2$, the same expression remains valid when one group size $n_i$ is continued to a positive real parameter $\lambda$ \cite{LSW2001b,LSW2002analyticity}.

No general exact expression of $\xi$ is known in three dimensions.
The only known exact result is the two-group case
$\xi(1,2)=\xi(2,1)=1$, first obtained in
Ref.~\cite{Lawler1989RandomSets}.
Beyond this special case, some general properties of the exponent spectrum have been established: as a function of the second group parameter $\lambda$, the exponent $\xi(k,\lambda)$ is strictly concave~\cite{Lawler1998StrictConcavity} and has recently been proved to be real analytic for $\lambda>0$~\cite{GaoEtAl2024}.

At the marginal dimension $d_c=4$, when each group contains only a single random walk, the asymptotic power law is replaced by a logarithmic form
\cite{Lawler1982FourDimensions,Lawler1985FourDimensions,Duplantier1988CMP},
\begin{equation}
    P_N(1,1,\ldots,1) \sim (\ln N)^{-p(p-1)/4}.
    \label{eq:four-dimensional-scaling}
\end{equation}
Using this logarithmic form, renormalization-group theory yields an $\epsilon$-expansion for the intersection exponent about $d=4-\epsilon$ \cite{Duplantier1988CMP},
\begin{equation}
\frac{\xi_{4-\epsilon}}{2} = \frac{p(p-1)}{8}\epsilon -\frac{p(p-1)(2p-5)}{32}\epsilon^2    +O(\epsilon^3).    \label{eq:epsilon-expansion}
\end{equation}
Setting $\epsilon=1$ formally gives an estimate for the three-dimensional case. For $p=2$, retaining the first two terms gives $\xi(1,1) = 5/8 = 0.625$, whereas numerical estimates give $\xi(1,1)\approx0.57$--$0.58$~\cite{BurdzyLawlerPolaski1989,LawlerVermesi2012}. The difference indicates that the low-order $\epsilon$-expansion provides only a rough estimate in three dimensions. Numerical simulations therefore provide an important approach for determining the exponents directly, yet systematic studies of three-dimensional intersection exponents remain limited.

In this paper, we use Monte Carlo (MC) simulations to determine intersection exponents for a series of cases in two and three dimensions. In two dimensions, the exponents obtained from our simulations are in good agreement with the exact results, providing a direct test of the accuracy
of our numerical approach. In three dimensions, we determine the intersection exponents $\xi(k,m)$ for two groups with $k=1,2$ and $m=1,2,\ldots,7$. We also obtain the exponents for several representative three-group cases. For the two-group case, we further continue the second group parameter from the integer $m$ to a positive real value $\lambda$ through moments of the conditional avoidance probability, and determine $\xi(k,\lambda)$ for $k=1,2,3$ and $0.25\leq\lambda\leq3$. Over the statistically resolved
ranges, these functions increase with $\lambda$ with decreasing slopes, consistent with the strict concavity of Brownian intersection exponents \cite{Lawler1998StrictConcavity}.

The remainder of this paper is organized as follows. Sec.~\ref{sec:methods} introduces the random-walk model, the integer path-group probabilities, and their continuous moment extension.  Sec.~\ref{sec:results} presents the corresponding results in two and three dimensions, together with the finite-size and sampling analyses. A summary and discussion are given in Sec.~\ref{sec:discussion}.

\section{Random-walk model and numerical methods}
\label{sec:methods}

\subsection{Integer numbers of random walks}

We simulate nearest-neighbor simple random walks $S_a(t)$ on $\mathbb Z^d$ in $d=2$ and $3$. At each step, a walker moves by one lattice spacing along one of the $2d$ coordinate directions with equal probability. The walks start from distinct sites on the boundary of a fixed square in two dimensions or a fixed cube in three dimensions, with side length $L_0$. Since $L_0$ remains fixed as $N$ increases, its value affects the amplitude and finite-$N$ corrections but not the asymptotic exponent.

For each realization, the walks are divided into $p$ groups containing $n_1,\ldots,n_p$ walks, and a separate hash table stores the sites visited by each group. All walks are advanced by one lattice spacing before the newly visited sites are checked against the hash tables of the other groups. A visit to a site already occupied by another group terminates the realization. Otherwise, the new sites are added to the corresponding hash tables and the walks continue to the next step. Since survival up to a given length implies survival at every shorter length, the same realization contributes to $P_N$ at all sampled values of $N$ reached before the first forbidden intersection. A realization without such an intersection is continued to $N_{\max}$ and counted as surviving at every sampled length.

For $B$ independent realizations, the non-intersection probability for any $N\leq N_{\max}$ is estimated as
\begin{equation}
    P_N=\frac{1}{B}\sum_{b=1}^{B}I_b(N) = \langle I_b(N)\rangle,     \label{eq:direct-estimator}
\end{equation}
where $I_b(N)=1$ if the $b$-th realization survives to $N$ and $I_b(N)=0$ otherwise. For a given $N$, $I_b(N)$ is a Bernoulli variable with success probability $P_N$. The statistical error of $P_N$ is therefore
\begin{equation}
 \delta(P_N) =\sqrt{\frac{P_N(1-P_N)}{B}}.  \label{eq:direct-error}
\end{equation}
It is pointed out that a realization that survives to step $N_1$ must also survive to $N_2<N_1$. Consequently, estimates of $P_N$ obtained at different walk lengths from the same realizations are correlated, although Eq.~(\ref{eq:direct-error}) gives the correct marginal error at each $N$.

\begin{table}[t]
\centering
\caption{Fitted values of the two-dimensional intersection exponent $\xi/2$. The fits of $P_N$ use Eq.~(\ref{eq:fit-ansatz}), and the exact values follow from Eq.~(\ref{eq:exact-2d-packets}). Entries labeled by $p$ correspond to $p$ groups with one walk in each group, i.e., $n_i=1$ for all $i=1,2,\ldots,p$.}
\label{tab:d2direct}
\begin{ruledtabular}
\begin{tabular}{c S[table-format=1.4(2)] S[table-format=1.6]}
{groups} & {MC} & {exact}\\
\hline
$p=2$ & 0.6242(5) & 0.625000\\
$p=3$ & 1.454(6) & 1.458333\\
$p=4$ & 2.63(4) & 2.625000\\
$p=5$ & 4.24(12) & 4.125000\\
\hline
$(1,2)$ & 0.998(3) & 1.000000\\
$(1,3)$ & 1.344(9) & 1.346500\\
$(1,4)$ & 1.68(2) & 1.678054\\
$(1,5)$ & 2.00(4) & 2.000000\\
$(1,6)$ & 2.30(5) & 2.315100\\
$(1,7)$ & 2.7(2) & 2.625000\\
$(2,2)$ & 1.47(2) & 1.458333\\
$(2,3)$ & 1.875(15) & 1.869167\\
$(2,4)$ & 2.25(3) & 2.255089\\
$(2,5)$ & 2.71(9) & 2.625000\\
$(2,6)$ & 2.94(8) & 2.983499\\
$(2,7)$ & 3.2(3) & 3.333333\\
\hline
$(1,1,2)$ & 2.02(4) & 2.000000\\
$(1,2,2)$ & 2.67(5) & 2.625000\\
$(2,2,2)$ & 3.35(9) & 3.333333\\
$(1,1,3)$ & 2.51(5) & 2.475167\\
\end{tabular}
\end{ruledtabular}
\end{table}

\begin{table}[t]
\centering
\caption{Fitted and exact values of the two-dimensional intersection exponent $\xi(k,\lambda)/2$ for $k=1,2,3$ and $0.25\leq\lambda\leq3$. The Monte Carlo values are obtained by fitting $P_N(k,\lambda)$ to Eq.~(\ref{eq:fit-ansatz}), and the exact values follow from Eq.~(\ref{eq:exact-2d}).}
\label{tab:d2}
\begin{ruledtabular}
\begin{tabular}{S[table-format=1.2]
                S[table-format=1.4(1)] S[table-format=1.4]
                S[table-format=1.3(2)] S[table-format=1.4]
                S[table-format=1.3(1)] S[table-format=1.4]}
{$\lambda$} & \multicolumn{2}{c}{$k=1$} & \multicolumn{2}{c}{$k=2$} & \multicolumn{2}{c}{$k=3$}\\
& {MC} & {exact} & {MC} & {exact} & {MC} & {exact}\\
\hline
0.25 & 0.328(6) & 0.2904 & 0.673(12) & 0.5673 & 1.04(3) & 0.8380\\
0.50 & 0.423(3) & 0.4128 & 0.770(8) & 0.7297 & 1.12(2) & 1.0314\\
0.75 & 0.5249(8) & 0.5224 & 0.879(4) & 0.8707 & 1.223(6) & 1.1966\\
1.00 & 0.6245(5) & 0.6250 & 0.998(2) & 1.0000 & 1.344(2) & 1.3465\\
1.25 & 0.7217(5) & 0.7230 & 1.117(2) & 1.1216 & 1.479(3) & 1.4864\\
1.50 & 0.8154(3) & 0.8177 & 1.232(2) & 1.2378 & 1.609(3) & 1.6191\\
1.75 & 0.9070(6) & 0.9098 & 1.343(2) & 1.3497 & 1.735(5) & 1.7463\\
2.00 & 0.9963(7) & 1.0000 & 1.451(2) & 1.4583 & 1.857(5) & 1.8692\\
2.25 & 1.0838(8) & 1.0885 & 1.555(3) & 1.5642 & 1.976(4) & 1.9884\\
2.50 & 1.1700(9) & 1.1756 & 1.655(4) & 1.6677 & 2.091(4) & 2.1046\\
2.75 & 1.254(2) & 1.2616 & 1.755(5) & 1.7693 & 2.202(5) & 2.2183\\
3.00 & 1.338(2) & 1.3465 & 1.853(6) & 1.8692 & 2.313(5) & 2.3297\\
\end{tabular}
\end{ruledtabular}
\end{table}

\begin{figure*}[t]
 \centering
 \includegraphics[width=\linewidth]{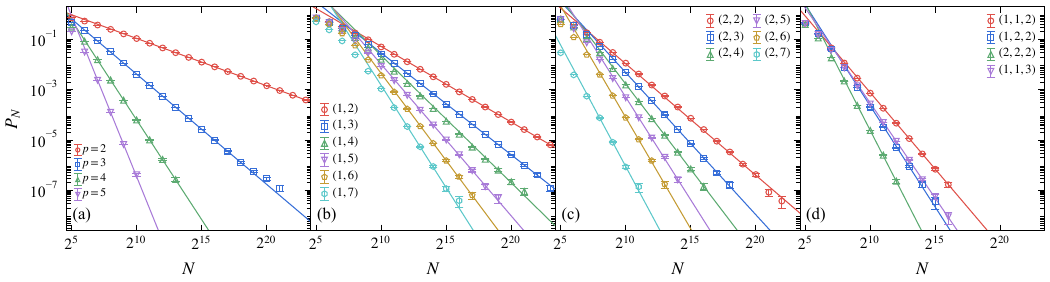}
 \caption{Non-intersection probability $P_N$ for integer path-group configurations of two-dimensional random walks. Panel (a) shows $p=2,\ldots,5$ groups with one walk in each group, i.e., $n_i=1$ for all $i=1,2,\ldots,p$. Panels (b) and (c) show $(1,m)$ and $(2,m)$, respectively, for $m=2,\ldots,7$. Panel (d) shows the three-group configurations $(1,1,2)$, $(1,2,2)$, $(2,2,2)$, and $(1,1,3)$. Open symbols are Monte Carlo data, and solid lines are the leading power laws obtained from fits to Eq.~(\ref{eq:fit-ansatz}).}
 \label{fig:direct2d}
\end{figure*}

\begin{figure}[!ht]
 \centering
 \includegraphics[width=\linewidth]{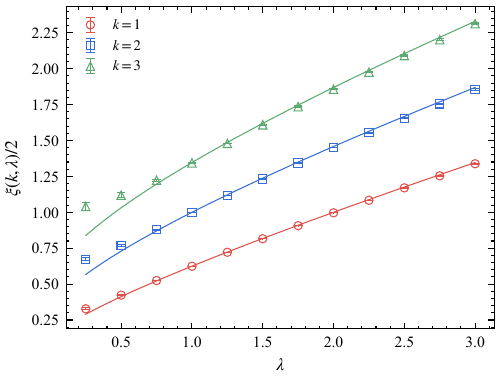}
 \caption{Two-dimensional intersection exponent $\xi(k,\lambda)/2$ for $k=1,2,3$ and $0.25\leq\lambda\leq3$. Open symbols are obtained from fits of $P_N(k,\lambda)$ to Eq.~(\ref{eq:fit-ansatz}), and solid curves show the exact result in Eq.~(\ref{eq:exact-2d}).}
 \label{fig:d2}
\end{figure}

\begin{figure*}[t]
 \centering
 \includegraphics[width=\linewidth]{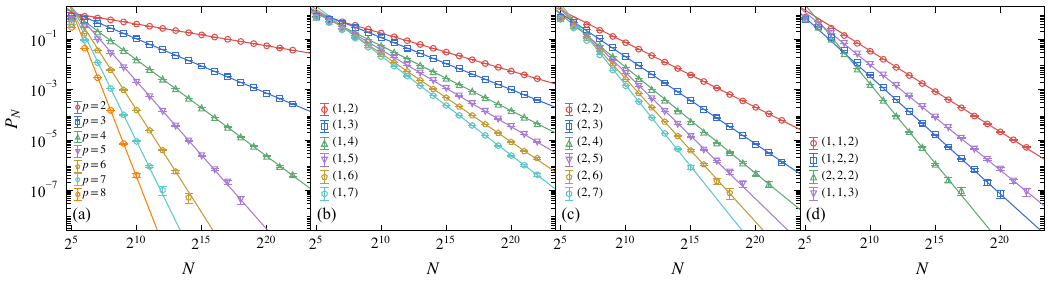}
 \caption{Non-intersection probability $P_N$ for integer path-group configurations of three-dimensional random walks. Panel (a) shows $p=2,\ldots,8$ groups with one walk in each group, i.e., $n_i=1$ for all $i=1,2,\ldots,p$. Panels (b) and (c) show $(1,m)$ and $(2,m)$, respectively, for $m=2,\ldots,7$. Panel (d) shows the three-group configurations $(1,1,2)$, $(1,2,2)$, $(2,2,2)$, and $(1,1,3)$. Open symbols are Monte Carlo data, and solid lines are the leading power laws obtained from fits to Eq.~(\ref{eq:fit-ansatz}).}
 \label{fig:direct}
\end{figure*}

\begin{figure*}[t]
 \centering
 \includegraphics[width=\textwidth]{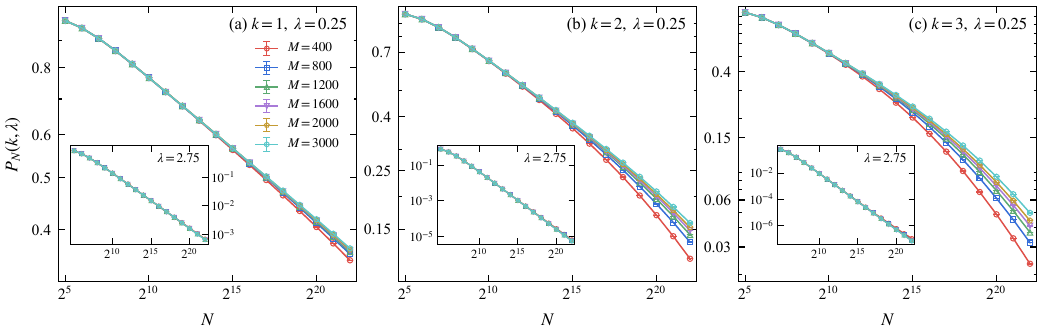}
 \caption{Dependence of the three-dimensional moment probability $P_N(k,\lambda)$ on the number $M$ of probe walks. The main panels show $\lambda=0.25$ for $k=1,2,$ and $3$, and the insets show the corresponding results at $\lambda=2.75$. Both the main panels and the insets use logarithmic axes. The variation between successive curves decreases with increasing $M$ and is weak at $M=2000$ over most of the sampled range.}
\label{fig:finiteM}
\end{figure*}

\subsection{Continuous extension of two-group exponents}

We next consider the two-group cases $(k,\lambda)$, where $\lambda$ may take
positive real values. The basic idea is to regard one group containing $k$
random walks as a fixed random environment and characterize the probability
that an additional random walk avoids this group
~\cite{Lawler1998StrictConcavity}.

For a given configuration of $k$ random walks, let $x_N$ denote the probability that an additional independent random walk does not intersect this group up to step $N$. If $m$ independent random walks form the second group, the conditional probability that all of them avoid the first group is $x_N^m$. Averaging over the configurations of the first group gives
\begin{equation}
 P_N(k,m)=\left\langle x_N^m\right\rangle.
\end{equation}
The second group size can therefore be continued from the positive integer $m$ to a positive real parameter $\lambda$ by defining
\begin{equation}
 P_N(k,\lambda)=\left\langle x_N^\lambda\right\rangle.
 \label{eq:moment-scaling}
\end{equation}
For integer $\lambda=m$, this expression reduces to the conventional two-group probability $P_N(k,m)$.

In the numerical calculation, however, the conditional probability $x_N$ is not known a priori. We therefore estimate it by generating $M$ independent random walks. If $M_0$ of them do not intersect the given $k$-walk group up to step $N$, we define the estimator
\begin{equation}
X_N=\frac{M_0}{M}.
\end{equation}
For a sufficiently large $M$, $X_N$ provides a binomial estimator of the fixed probability $x_N$. The numerical estimate of Eq.~(\ref{eq:moment-scaling}) is then obtained by replacing $x_N^\lambda$ with $X_N^\lambda$ for each configuration of the $k$-walk group and averaging over independent configurations.

For finite $M$, although $X_N$ is an unbiased estimator of $x_N$, its nonlinear powers are generally biased. To quantify this effect, write
\begin{equation}
X_N = x_N + \delta,
\end{equation}
where $\delta$ denotes the statistical deviation of the binomial estimator. Expanding $X_N^\lambda$ around $x_N$ gives
\begin{equation}
X_N^\lambda = x_N^\lambda + \lambda x_N^{\lambda-1}\delta +\frac{1}{2}\lambda(\lambda-1) x_N^{\lambda-2} \delta^2 +O(\delta^3).  \label{eq-xnl}
\end{equation}
Averaging over the binomial sampling gives
\begin{equation}
\langle \delta \rangle=0,
\qquad
\langle \delta^2\rangle =\frac{x_N(1-x_N)}{M}.       \label{eq-delta}
\end{equation}
Thus, averaging Eq.~(\ref{eq-xnl}), we obtain
\begin{equation}
\left\langle X_N^\lambda\right\rangle
= x_N^\lambda + \frac{\lambda(\lambda-1)}{2M} x_N^{\lambda-1}(1-x_N) + \cdots.
\label{eq:finite-m-convexity}
\end{equation}

For $\lambda=1$, $\langle X_N\rangle=x_N$ is unbiased. For $0<\lambda<1$, the leading finite-$M$ correction is negative, whereas for $\lambda>1$ it is positive. This finite-$M$ bias becomes significant when $x_N$ is small (large $N$) and should be taken into account when interpreting the numerical estimates of the intersection exponent. In particular, for $0<\lambda<1$, the negative bias increasingly underestimates $P_N(k,\lambda) = \langle X_N^\lambda\rangle$ as $N$ increases, leading to an overestimate of $\xi$. For $\lambda>1$, the bias has the opposite sign and leads to an underestimate of $\xi$. This characteristic of the numerical scheme will be verified in the simulations.

\section{Numerical results}
\label{sec:results}

To determine the exponents, we fit the probability $P_N$ to the scaling ansatz
\begin{equation}
 P_N=N^{-\xi/2}\left(a_0+a_1N^{-\omega_1}+a_2N^{-\omega_2}\right), \label{eq:fit-ansatz}
\end{equation}
One or both correction terms are omitted when they are not required by the data.  We apply a lower cutoff $N\geq N_{\min}$ and accept fits with $\chi^2/\mathrm{DF}\simeq1$, where DF denotes the number of degrees of freedom.  For the integer-group data, the values of $P_N$ at different $N$ are correlated because the same realizations contribute at several walk lengths. A fit that treats these values as independent counts the same histories more than once and can underestimate the uncertainty of the exponent. The final estimates are obtained from the stable fits under changes of $N_{\min}$, the number of correction terms, and the correction exponents $\omega_i$.

\subsection{Intersection exponents in two dimensions}
\label{sec:d2}

The exact two-dimensional exponents provide a direct test of the numerical calculation before we turn to three dimensions. Figure~\ref{fig:direct2d} shows $P_N$ for the integer path groups studied here. The probabilities follow power laws over the resolved ranges. The data reach $N=2^{23}$ for the slowly decaying cases, while the largest length for $p=5$ groups with $n_i=1$ is $N=2^{10}$. The shorter range in the latter case results from the rapid decrease in the number of surviving realizations. Fits to Eq.~(\ref{eq:fit-ansatz}) give the exponents listed in Table~\ref{tab:d2direct}.

The estimates in Table~\ref{tab:d2direct} closely follow Eq.~(\ref{eq:exact-2d-packets}) for complete mutual, two-group, and three-group avoidance. This agreement over configurations with substantially different decay rates shows that the leading exponents can be separated from the finite-size corrections over the selected ranges. As the exponent increases, the mean number $B P_N$ of surviving samples decreases and the uncertainty grows, in accordance with Eq.~(\ref{eq:direct-error}).

We next turn to the continuous moment order.  In two dimensions, the exact result is \cite{LSW2002analyticity}
\begin{equation}
\xi(k,\lambda) =\frac{\left(\sqrt{24k+1}+\sqrt{24\lambda+1}-2\right)^2-4}{48}.
 \label{eq:exact-2d}
\end{equation}
The results in Fig.~\ref{fig:d2} and Table~\ref{tab:d2} were obtained with $M=2000$ probe walks for each background configuration. At $\lambda=1$, the estimates $0.6245(5)$, $0.998(2)$, and $1.344(2)$ agree with the exact values $0.625$, $1$, and $1.346500\ldots$. They also agree with the corresponding integer path-group estimates, as required by Eq.~(\ref{eq:moment-scaling}).

Away from $\lambda=1$, the comparison with Eq.~(\ref{eq:exact-2d}) displays the finite-$M$ effects described by Eq.~(\ref{eq:finite-m-convexity}). For $\lambda<1$, configurations with small $x_N$ are increasingly underrepresented at large $N$. The measured probability then decreases too rapidly, shifting $\xi(k,\lambda)$ above the exact value.
For $\lambda>1$, the measured moments are enhanced more strongly at large $N$, and the fitted exponents generally lie below the exact values. The two-dimensional comparison therefore tests the moment calculation and identifies the loss of accuracy at the lowest moment order before the method is applied in three dimensions, where no exact function is available.

\subsection{Intersection exponents in three dimensions}
\label{sec:d3}

In Fig.~\ref{fig:direct}, we show $P_N$ for integer group sizes in three dimensions. The probabilities $P_N$ follow power laws throughout the resolved ranges, indicating a well-defined intersection exponent $\xi$. Fits to Eq.~(\ref{eq:fit-ansatz}) give the exponents listed in Table~\ref{tab:direct}. The $(1,m)$ data reach $N_{\max}=2^{23}$ for all $m$, while the maximum step is $N_{\max}=2^{10}$ for eight mutually avoiding walks and $N_{\max}=2^{16}$ for the $(2,7)$ configuration.

Let us first compare our results with existing results.
The only exactly known exponent in three dimensions is
$\xi(1,2)=\xi(2,1)=1$~\cite{Lawler1989RandomSets}.
Our fit gives $\xi(1,2)/2=0.5004(4)$, in agreement with this exact result. For $\xi(1,1)/2$, an earlier study obtained a value near $0.29$~\cite{BurdzyLawlerPolaski1989}, while a more extensive simulation gave a value close to $0.285$~\cite{LawlerVermesi2012}. Our result, $\xi(1,1)/2=0.2872(2)$, is consistent with both estimates and provides a more precise numerical value. These comparisons support the reliability of our numerical method.

Our calculations also cover a much wider range of group sizes than the existing exact results. As listed in Table~\ref{tab:direct}, $\xi(k,m)/2$ for the $(1,m)$ sequence increases from $0.5004(4)$ at $m=2$ to $1.31(1)$ at $m=7$, while the $(2,m)$ sequence increases from $0.8553(12)$ to $2.14(4)$. Increasing the number of walks in the second group imposes more avoidance constraints, causing $P_N$ to decay faster and hence increasing $\xi$. For complete mutual avoidance, where each walk forms a separate group, $\xi/2$ increases from $0.2872(2)$ for two walks to $4.4(1)$ for eight walks.

For a fixed total number of walks, $\xi/2$ increases as the number of groups increases, since more intersections are forbidden. For four walks, the results are $0.8553(12)$ for $(2,2)$, $1.069(3)$ for $(1,1,2)$, and $1.271(5)$ for $(1,1,1,1)$. For five walks, the corresponding values are $1.151(7)$ for $(2,3)$, $1.388(4)$ for $(1,1,3)$, and $1.95(3)$ for $(1,1,1,1,1)$. Thus, allowing intersections within larger groups systematically lowers the exponent.

Two pairs of configurations with different groupings also yield nearly equal exponents. The difference between $(1,4)$ and $(2,2)$ is $0.0013(14)$, while that between $(2,4)$ and $(1,1,3)$ is $0.01(2)$. Both differences are consistent with zero within the quoted uncertainties. This numerical agreement suggests a possible geometric relation between the corresponding avoidance constraints in three dimensions. Whether either relation is exact remains an open question.

\begin{table}[t]
\centering
\caption{Fitted values of the three-dimensional intersection exponent $\xi/2$. The fits of $P_N$ use Eq.~(\ref{eq:fit-ansatz}). Entries labeled by $p$ correspond to $p$ groups with one walk in each group, i.e., $n_i=1$ for all $i=1,2,\ldots,p$.}
\label{tab:direct}
\setlength{\tabcolsep}{3pt}
\begin{ruledtabular}
\begin{tabular}{c S[table-format=1.4(2)] c S[table-format=1.4(2)] c S[table-format=1.3(2)]}
{groups} & {MC} & {groups} & {MC} & {groups} & {MC}\\
\hline
$p=2$ & 0.2872(2) & $(1,2)$ & 0.5004(4) & $(1,1,2)$ & 1.069(3)\\
$p=3$ & 0.7169(4) & $(1,3)$ & 0.6858(4) & $(1,2,2)$ & 1.55(2)\\
$p=4$ & 1.271(5) & $(1,4)$ & 0.8566(6) & $(2,2,2)$ & 2.09(3)\\
$p=5$ & 1.95(3) & $(1,5)$ & 1.014(2) & $(1,1,3)$ & 1.388(4)\\
$p=6$ & 2.70(6) & $(1,6)$ & 1.166(8) & &\\
$p=7$ & 3.5(2) & $(1,7)$ & 1.31(1) & &\\
$p=8$ & 4.4(1) & $(2,2)$ & 0.8553(12) & &\\
& & $(2,3)$ & 1.151(7) & &\\
& & $(2,4)$ & 1.40(2) & &\\
& & $(2,5)$ & 1.666(8) & &\\
& & $(2,6)$ & 1.89(2) & &\\
& & $(2,7)$ & 2.14(4) & &\\
\end{tabular}
\end{ruledtabular}
\end{table}

\begin{table}[t]
\centering
\caption{Fitted values of the three-dimensional intersection exponent $\xi(k,\lambda)/2$ for $k=1,2,3$ and $0.25\leq\lambda\leq3$. The fits of $P_N(k,\lambda)$ use Eq.~(\ref{eq:fit-ansatz}).}
\label{tab:d3}
\begin{ruledtabular}
\begin{tabular}{S[table-format=1.2] S[table-format=1.5(2)] S[table-format=1.4(2)] S[table-format=1.3(2)]}
{$\lambda$} & {$k=1$} & {$k=2$} & {$k=3$}\\
\hline
0.25 & 0.0915(9) & 0.237(13) & 0.488(13)\\
0.50 & 0.1610(3) & 0.317(8) & 0.535(14)\\
0.75 & 0.2264(2) & 0.404(3) & 0.598(11)\\
1.00 & 0.2871(2) & 0.5004(3) & 0.6852(4)\\
1.25 & 0.3438(4) & 0.5948(6) & 0.790(7)\\
1.50 & 0.3976(4) & 0.6841(9) & 0.898(11)\\
1.75 & 0.4491(7) & 0.7700(5) & 1.00(2)\\
2.00 & 0.4988(9) & 0.8508(5) & 1.112(9)\\
2.25 & 0.5465(12) & 0.9281(5) & 1.215(8)\\
2.50 & 0.593(1) & 1.0025(6) & 1.313(7)\\
2.75 & 0.637(2) & 1.0743(7) & 1.406(7)\\
3.00 & 0.681(2) & 1.1438(7) & 1.496(6)\\
\end{tabular}
\end{ruledtabular}
\end{table}

We next consider the extension to noninteger $\lambda$. Since no exact expression is available for the continuous three-dimensional exponents, we first examine the dependence of $P_N(k,\lambda)$ on the number $M$ of probe walks. Figure~\ref{fig:finiteM} shows the results at $\lambda=0.25$ and $2.75$ for $k=1,2,$ and $3$.

The curves approach a stable range as $M$ increases. At $M=2000$, the remaining dependence on $M$ is weak over most of the values of $N$, $k$, and $\lambda$ shown in Fig.~\ref{fig:finiteM}, although it remains visible at the largest $N$ for $k=3$ and $\lambda=0.25$. The computational time of the inner sampling increases approximately in proportion to $M$, while increasing $M$ beyond this range produces little change in most of the measured probabilities. We therefore use $M=2000$ for the three-dimensional continuous-moment calculation and retain the low-$\lambda$ limitation identified in the two-dimensional comparison.

\begin{figure}[t]
 \centering
 \includegraphics[width=\linewidth]{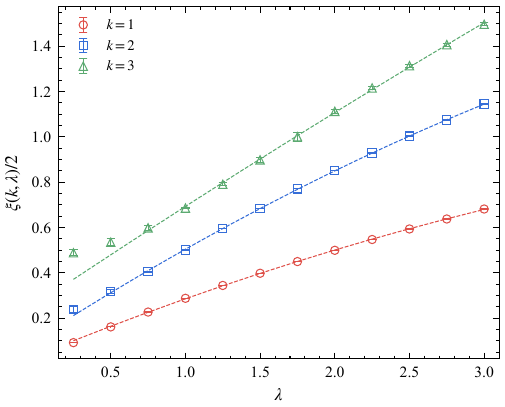}
 \caption{Three-dimensional intersection exponent $\xi(k,\lambda)/2$ for $k=1,2,3$ and continuous moment order $0.25\leq\lambda\leq3$. Open symbols are obtained from fits to Eq.~(\ref{eq:fit-ansatz}), and dashed lines are guides to the eye. No exact dependence on $\lambda$ is known in three dimensions.}
 \label{fig:d3}
\end{figure}

With this value of $M$, Fig.~\ref{fig:d3} and Table~\ref{tab:d3} show $\xi(k,\lambda)/2$ for $k=1,2,3$ and $0.25\leq\lambda\leq3$. The exponent increases with both $\lambda$ and $k$. Strict concavity in $\lambda$ is known for the Brownian intersection exponent in three dimensions \cite{Lawler1998StrictConcavity}. The estimates away from the lowest moment orders increase with a decreasing slope and are consistent with this property. The points at $\lambda=0.25$, particularly for $k=2$ and $3$, are affected by the finite-$M$ bias identified in two dimensions and are not used to infer the shape of the continuous dependence. No exact function is available in three dimensions, and the integer exponents provide reference values only at selected moment orders.

At $\lambda=1$, the moment is linear in the conditional probability and is unbiased for every $M$.  The estimates for $k=1,2,3$ are $0.2871(2)$, $0.5004(3)$, and $0.6852(4)$, in agreement with the corresponding integer results $0.2872(2)$, $0.5004(4)$, and $0.6858(4)$.  This agreement provides a direct check of Eq.~(\ref{eq:moment-scaling}) at the point where the finite-$M$ estimator is unbiased.

\section{Discussion}
\label{sec:discussion}

We have studied the non-intersection probability of multiple simple random
walks in two and three dimensions, with intersections allowed within each
group but forbidden between different groups. In two dimensions, our
numerical results agree with the exact intersection exponents for a range
of group configurations, validating the numerical procedure. In three
dimensions, we determine $\xi(k,m)$ for two groups with $k=1,2$ and
$m=1,2,\ldots,7$, together with several representative three-group cases.
We further continue the second group parameter to a positive real value
$\lambda$, obtaining $\xi(k,\lambda)$ for $k=1,2,3$ and
$0.25\leq\lambda\leq3$. Away from the lowest moment orders, these functions
increase with $\lambda$ with decreasing slopes, in agreement with the strict concavity of the Brownian intersection exponent \cite{Lawler1998StrictConcavity}.

No explicit expression for $\xi$ is currently known in three dimensions. The
present results provide quantitative benchmarks for future probabilistic
and fixed-dimensional field-theory calculations, including the dependence
on the number and grouping of paths, the values at noninteger $\lambda$,
and the variation of the slope across the same parameter range.

Several questions remain open. Rare-event sampling and an adaptive choice
of the number of inner samples could extend the numerical study to larger
groups and a wider range of $\lambda$. The close values of the exponents
for $(1,4)$ and $(2,2)$, and for $(2,4)$ and $(1,1,3)$, also suggest possible
relations between different group configurations. Whether these
similarities follow from exact relations between the corresponding
non-intersection constraints remains an open question.

\section*{Acknowledgments}
Q.S. and Y.D. were supported by the National Natural Science Foundation of China under
Grant No.~12275263 and the Quantum Science and Technology National Science and Technology
Major Project under Grant No.~2021ZD0301900. M.L. was supported by the National Natural
Science Foundation of China under Grant No.~12675037. R.L. and X.L. were supported by the
National Key R\&D Program of China under Grant No.~2021YFA1002700 and the Beijing Natural
Science Foundation under Grant No.~JQ26001.

\bibliography{references}

\end{document}